\pdfoutput=1
\documentclass[sigconf,nonacm]{acmart}

\usepackage{amsmath}
\usepackage{array}
\usepackage{booktabs}
\usepackage{graphicx}
\usepackage{multirow}
\usepackage{xcolor}
\usepackage{balance}

\setcopyright{none}
\renewcommand\footnotetextcopyrightpermission[1]{}
\hypersetup{colorlinks=true,linkcolor=black,citecolor=black,urlcolor=blue}
\graphicspath{{figures/}}

\title[Predict Before Replay]{Predict Before Replay: Joint FEC and Flight Control for Reliable Scale-Up Links}

\author{Fan Yang}
\affiliation{%
  \institution{Institute of Computing Technology, Chinese Academy of Sciences}
  \city{Beijing}
  \country{China}}
\email{yangfan2020@ict.ac.cn}

\author{Jiaqi Liu}
\affiliation{%
  \institution{Institute of Computing Technology, Chinese Academy of Sciences}
  \city{Beijing}
  \country{China}}
\affiliation{%
  \institution{Hangzhou Institute for Advanced Study, UCAS}
  \city{Hangzhou}
  \country{China}}
\email{liujiaqi242@mails.ucas.ac.cn}

\author{Tao Jiang}
\affiliation{%
  \institution{Institute of Computing Technology, Chinese Academy of Sciences}
  \city{Beijing}
  \country{China}}
\email{jt@ncic.ac.cn}

\author{Zhan Wang}
\affiliation{%
  \institution{Institute of Computing Technology, Chinese Academy of Sciences}
  \city{Beijing}
  \country{China}}
\email{wangzhan@ncic.ac.cn}

\begin{document}
\raggedbottom

\begin{abstract}
Scale-up accelerator fabrics send latency-sensitive flits over serial links at
hundreds of gigabits per second.  Their reliability pipeline first relies on
FEC, then detects residual failures and replays unacknowledged data.  At these
line rates, delayed feedback lets later flits enter the replay window before a
residual failure is reported, so standard replay can amplify one corrupted flit
into a suffix retransmission.  This paper presents \emph{PREFACE}, a pre-FEC
controller for temporally correlated burst errors.  A two-state Bayesian filter
converts corrected-symbol observations into a next-flit burst posterior and
jointly selects FEC strength with an outstanding-flit cap.  We implement
PREFACE in ns-3 with publicly verifiable UALink 200G 1.0 replay semantics.
PREFACE improves goodput by 10.52\%, lowers P99 latency by 50.75\%, cuts
replay by 47.52\%, and improves modeled ring
AllReduce by 13.1--27.0\%.
\end{abstract}

\maketitle

\section{Introduction}

Scale-up networks connect GPUs, accelerators, and memory devices over links
whose latency is short enough that link-local recovery is attractive, yet whose
line rate makes every nanosecond of delayed reaction consequential
\cite{ieee8023ck,oifcei,nvidiah100}.  UALink, Scale Up Ethernet,
CXL-derived proposals, proprietary accelerator links, and PCIe 6.0-style
high-speed I/O all combine error detection, FEC, and some form of replay or
retry \cite{ualink10,sue,rxl,pcie60}.  These mechanisms are usually studied as separate
choices: spend redundancy continuously, or detect an error and retransmit.

Strong FEC is highly effective for sparse, approximately independent raw
errors.  PREFACE targets a different regime: short, temporally correlated
bursts in which adjacent flits see elevated symbol-error counts before feedback
returns.  Such bursts can arise from decision-feedback equalizer (DFE) error
propagation, crosstalk, transient signal-integrity loss, or equalizer
instability \cite{dfe}.  The remaining challenge is limiting the performance
cost when a residual error still activates standard replay.

The separation hides a feedback-amplification problem.  Suppose a transmitter
can retain $W$ flits and a burst corrupts flit $i$.  Until a replay request
returns, later flits continue to enter the replay buffer.  With standard replay,
recovery begins at $i$ and retransmits the retained suffix.  Thus the cost of a
residual error depends not only on FEC strength but also on how many flits were
admitted while the link was likely bad.  A controller that changes coding but
leaves exposure unchanged can still create a large replay episode; a controller
that waits for an uncorrectable decode is necessarily late.

Modern receivers already observe corrected errors before those errors become
replay events.  Error counts and FEC statistics are especially informative for
burst channels caused by crosstalk, equalizer error propagation, or transient
signal degradation \cite{dfe}.  The opportunity is to use that early signal for
both redundancy and admission, while leaving standard recovery intact.

We present \emph{PREFACE}: \textbf{pre}-\textbf{FEC} adaptive coding and
exposure control.  PREFACE tracks a two-state burst posterior from pre-FEC
symbol-error observations.  For each next flit, it compares three coding
profiles by balancing the redundancy they add against the expected replay cost
of an uncorrected residual error.  When the
posterior selects strong protection, the controller also caps outstanding
traffic by an absolute risk budget.  A single burst-risk estimate therefore
coordinates both decisions: which FEC profile to use, and how many flits may
remain unacknowledged during risky intervals.

This paper makes four contributions:

\begin{itemize}
  \item It identifies \emph{replay exposure} as the missing coupling between
  adaptive FEC and delayed standard replay in high-rate scale-up links.
  \item It designs PREFACE, a cost-aware burst predictor that jointly controls
  FEC strength and outstanding flight without changing replay correctness.
  \item It implements a reusable ns-3 model with UALink 200G 1.0-inspired
  standard-replay semantics and serialized reverse control, and evaluates fixed
  FEC, post-decode adaptation, predictive-FEC, and joint-control baselines.
  \item It shows statistically significant link-level and modeled ring
  AllReduce gains, a strict joint-control ablation, sensitivity to feedback and
  buffer size, and a long-run reliability invariant check.
\end{itemize}

\section{Background and Motivation}

\subsection{FEC, replay, and burst errors}

FEC converts a bounded number of corrupt symbols into extra wire bytes and
decode latency \cite{reedsolomon,ieee8023bs,pcie60fec}.  Replay spends bandwidth
only after residual corruption, but each replay event also waits for feedback
and blocks later payload flits while the retained suffix is retransmitted.
RIFL demonstrates efficient link-local recovery in short data-center links with
NACK-only, in-band retransmission \cite{rifl}.  UALink-style scale-up links
expose a different replay path, with retained transmit state and standard replay
behavior \cite{ualink10}; PREFACE evaluates predictive control on this
UALink-style model.

Independent bit errors are an incomplete stress model for a high-speed
electrical link.  Decision-feedback equalizer propagation and related channel
effects can create multi-symbol bursts \cite{dfe}.  We therefore use the
Gilbert--Elliott two-state abstraction \cite{gilbert,elliott}: a mostly clean
GOOD state and a persistent, high-error BAD state.  This controlled burst model
makes the mechanism stress explicit; Section~\ref{sec:validity} outlines how
measured pre-FEC traces can calibrate the parameters for particular cables,
connectors, and post-FEC reliability targets.

Table~\ref{tab:positioning} makes the problem boundary explicit.  PREFACE targets
short correlated bursts where corrected-symbol evidence arrives early enough to
change both coding and outstanding flight before delayed replay feedback returns.

\begin{table}[t]
\centering
\caption{Mechanism positioning by error regime.}
\label{tab:positioning}
\resizebox{\columnwidth}{!}{%
\renewcommand{\arraystretch}{1.22}%
\begin{tabular}{>{\raggedright\arraybackslash}p{0.25\columnwidth}>{\raggedright\arraybackslash}p{0.34\columnwidth}>{\raggedright\arraybackslash}p{0.31\columnwidth}}
\toprule
Approach & Problem setting & Exposure implication \\
\midrule
\textbf{1) Fixed strong FEC} & Sparse independent raw errors; always-on protection & Lowers residual probability; retained suffix size unchanged \\
\textbf{2) Post-decode adapt.} & Residual failure visible after FEC/CRC & Reacts after a replay request and suffix have formed \\
\textbf{3) Predictive FEC} & Corrected-symbol evidence before failure & Raises protection early; leaves outstanding flight unchanged \\
\textbf{4) Link-local retry} & Detected link loss with NACK/retry path & Optimizes retry semantics \\
\textbf{5) PREFACE} & Short correlated bursts with early pre-FEC evidence & Raises protection and caps flight before feedback returns \\
\bottomrule
\end{tabular}}
\end{table}

\subsection{Why post-decode adaptation is late}

Let a 640-byte flit traverse a 400-Gb/s link.  Its payload serialization alone
is 12.8~ns.  A 50-ns request delay plus propagation spans multiple new flit
admissions.  A post-decode controller updates only when a codeword is declared
uncorrectable.  By then, (1) a replay request is inevitable, (2) later flits may
occupy TxReplay, and (3) standard replay will resend a suffix rather than only
the damaged flit.

Always-on strong FEC avoids many such events, but its 6.25\% modeled wire
overhead and 12-ns decode delay apply even in the GOOD state.  Predictive FEC can
remove unnecessary redundancy; however, if it keeps the same $W$ during a BAD
posterior, the retained suffix remains exposed to a rare residual failure.
This motivates a joint action:

\begin{quote}
Use corrected-error evidence before replay, and spend both coding and flight
budget according to the predicted cost of the next residual failure.
\end{quote}

\subsection{Replay exposure as an amplification term}

The key quantity couples the probability of an uncorrectable flit with
the amount of still-valid traffic that becomes coupled to it.  Let $i$ be the
first failed sequence, $O_i$ the outstanding set when its replay request is
processed, and $S_i=|\{j\in O_i:j\ge i\}|$ the retained suffix.  Standard
replay sends at least $S_i$ payload flits for that episode.  If $T_f$ is the
flit admission interval and $T_d$ is decode plus request-return time, the
number exposed before feedback is bounded approximately by

\begin{equation}
E_i \le \min\!\left(W,\;O_i^{0}+\left\lceil T_d/T_f\right\rceil\right),
\label{eq:exposure-bound}
\end{equation}

where $O_i^{0}$ is occupancy at failure.  The exact suffix also depends on ACK
progress and request serialization, but Eq.~\eqref{eq:exposure-bound} exposes
the two control levers: FEC reduces the probability that an episode begins,
while a flight cap limits its conditional size.  For coding action $k$ and
window $W$, a first-order expected wire-cost decomposition is

\begin{equation}
C(k,W\mid q) \approx o_k + r_k(q)\,\mathrm{E}[S\mid W,q] + u(W),
\label{eq:joint-cost}
\end{equation}

where $o_k$ is redundancy, $r_k$ residual failure probability, and $u(W)$ the
underfill cost of constraining flight.  Optimizing $k$ while holding $W$ fixed
can reduce $r_k$ yet leave a large conditional replay suffix.  Conversely,
constraining $W$ without early evidence pays $u(W)$ even while the channel is
healthy.  PREFACE uses the same posterior to approximate this joint decision,
but deliberately decomposes it into a coding selector and a two-level flight
budget so that the datapath remains simple.

\section{PREFACE Design}

\begin{figure*}[t]
  \centering
  \includegraphics[width=0.96\textwidth]{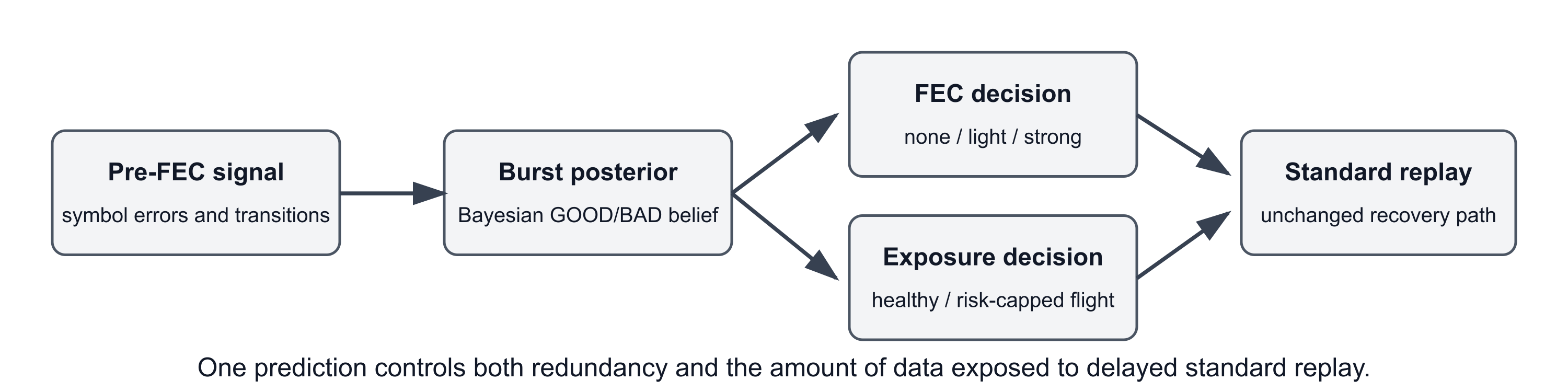}
  \caption{PREFACE uses one pre-FEC burst posterior for two coupled decisions.
  Standard replay remains the correctness backstop.}
  \label{fig:mechanism}
\end{figure*}

\subsection{Burst posterior}

PREFACE needs an early link-health signal before a decode failure.  A protected
profile can expose a corrected-symbol record or coarser high-SER indication
without requiring a full raw-bit trace.  A pure FEC-off mode exposes CRC failure
after the prediction window has passed.  We therefore describe the weakest
deployable profile as LOW: a minimal monitor/protection mode, or an equivalent
PHY-side pre-FEC telemetry path.  In the simulator, a threshold model supplies
synthetic pre-FEC symbol counts to predictive policies; Section~\ref{sec:validity}
describes how measured telemetry can replace this signal.

For the controller, let $z_t\in\{0,1\}$ indicate whether the just-received flit
contained any raw symbol error, including errors corrected by FEC.  Let
$q_t^{-}$ be the prior probability that flit $t$ is in the BAD state, and let
$e_G$ and $e_B$ be the corruption emission probabilities.  The posterior is

\begin{equation}
q_t=\frac{q_t^{-}P(z_t\mid B)}
{q_t^{-}P(z_t\mid B)+(1-q_t^{-})P(z_t\mid G)}.
\end{equation}

With GOOD-to-BAD transition probability $a$ and BAD-to-GOOD probability $b$,
the next-flit prior becomes

\begin{equation}
q_{t+1}^{-}=q_t(1-b)+(1-q_t)a.
\label{eq:predict}
\end{equation}

The primary evaluation fixes $(a,b,e_G,e_B)$ before held-out testing.  The code
also contains an online expected-count estimator for future trace-driven
extensions.  The fixed binary emission model keeps the mechanism results
independent of any one decoder's analog confidence scale.

\subsection{Cost-aware FEC selection}

The coding menu contains LOW, LIGHT, and STRONG profiles.  Profile $k$ has
normalized overhead $o_k$ and corrects up to $c_k$ symbols.  STRONG is modeled
after a KP4-like RS(544,514) correction strength: 30 parity symbols correct up
to 15 symbol errors \cite{reedsolomon,ieee8023bs,g7095}.  LIGHT is modeled as a
lower-overhead shortened or monitoring profile rather than a standardized
Ethernet FEC mode.  The evaluation uses
profile-level correction thresholds, byte overheads, and decode delays rather
than a bit-accurate RS/BCH decoder.

The simulator's synthetic emission model bounds error magnitude by $m_G$ or
$m_B$ in each state.  PREFACE estimates residual failure as

\begin{align}
r_k(q)={}&(1-q)e_G\left[1-\frac{c_k}{m_G}\right]^+ \\
         &+q e_B\left[1-\frac{c_k}{m_B}\right]^+,
\end{align}

where $[x]^+=\max(0,x)$.  It selects

\begin{equation}
k^*=\arg\min_k\{o_k+\lambda r_k(q_{t+1}^{-})\}.
\label{eq:cost}
\end{equation}

$\lambda$ represents the normalized cost of a residual failure under standard
replay.  Values are swept on training seeds 1--10, and $\lambda=16$ is frozen
before all reported seeds 101 and above.  The estimator is a simple control
heuristic for converting bounded symbol magnitudes into residual-risk estimates.

\subsection{Exposure control}

Let $W_H$ be the healthy TxReplay/admission window and $W_R$ an absolute risk
budget.  PREFACE sets

\begin{equation}
W_t=\begin{cases}
\min(W_H,W_R), & k^*=\mathrm{STRONG},\\
W_H, & \text{otherwise}.
\end{cases}
\label{eq:window}
\end{equation}

The primary point uses $W_H=16$ and $W_R=8$.  The healthy window is a baseline
admission-capacity setting chosen for the modeled link.  A 640-byte flit serializes in
12.8~ns at 400~Gb/s, while ACK return spans tens to roughly one hundred
nanoseconds.  Sixteen flits avoid obvious ACK-limited underfill and define
about 10~KB of retained-suffix exposure; the risk cap halves this to about
5~KB.

A smaller window limits suffix amplification and payload-blocking exposure, but an
aggressive contraction can underfill the link.  We choose $W_R$ as an absolute
feedback-delay/burst budget rather than as a fixed offset from $W_H$.  It is selected on
training seeds together with $\lambda$ and frozen before the primary,
robustness, and collective tests.  The optimal risk budget can change with
burst persistence, load, feedback delay, and FEC costs; online adaptation is
future work.

Exposure control changes performance, not correctness.  ACK processing,
sequence ordering, retry limits, and standard replay are unchanged.  If the
predictor is wrong, the link may spend excess FEC or lose throughput; detected
residual errors still follow the baseline replay path.

\subsection{Implementation considerations}

The required state is one posterior, four model probabilities, three profile
costs, and a current admission limit.  Updates occur once per decoded flit and
use additions, multiplications, and comparisons; the arithmetic is amenable to a
fixed-point datapath.  Pre-FEC error counts may be exported from an FEC decoder
locally, avoiding a new wire protocol.  Profile signaling and safe transition
epochs are implementation-specific; the simulator represents them as immediate
selection.  Real hardware must synchronize TX/RX profile state, protect profile
metadata, and define replay across profile boundaries.

\subsection{Controller operation and safety envelope}

For each decoded flit, the receiver forms $z_t$ from the pre-FEC correction
record, updates Eq.~\eqref{eq:predict}, evaluates the three costs in
Eq.~\eqref{eq:cost}, and publishes the selected profile and risk state.  The
transmit side applies Eq.~\eqref{eq:window} to new admission.  Existing replay
entries are never discarded when the budget contracts; the controller simply
blocks new payload until occupancy falls below $W_t$.  This distinction is
essential because forcibly trimming TxReplay would change correctness.

\begin{figure}[t]
\centering
\fbox{\begin{minipage}{0.93\columnwidth}
\small\ttfamily
on decoded flit $t$ with pre-FEC observation $z_t$:\\
\quad 1. update posterior $q_t$ and predict $q^-_{t+1}$\\
\quad 2. for $k\in\{$LOW,LIGHT,STRONG$\}$ compute $o_k+\lambda r_k$\\
\quad 3. select minimum-cost profile $k^*$\\
\quad 4. set budget $W_t\leftarrow W_R$ iff $k^*=$STRONG\\
\quad 5. admit only if replay occupancy $<W_t$\\
on detected residual error: execute unchanged standard replay
\end{minipage}}
\caption{PREFACE controller pseudocode.  Budget contraction blocks admission;
it never invalidates retained state.}
\label{fig:pseudocode}
\end{figure}

False positives spend strong FEC and may temporarily block admission.  False
negatives fall back to the standard retry path.  Hence prediction quality
affects performance, while detected-error correctness remains governed by the
underlying replay protocol.  A deployment can also pin $W_t=W_H$ or force
STRONG FEC as a fail-safe profile if telemetry is absent, the posterior becomes
invalid, or a profile transition cannot be completed safely.

\section{Methodology}

\subsection{ns-3 link model}

We add a \texttt{scaleup-reliability} contribution to the ns-3 simulator
\cite{ns3}.  Each link contains a serializer, propagation and decode delay,
bounded TxReplay state, an in-order receive buffer, a reverse-control serializer,
and a seeded burst-error model.  A failed decode schedules a request; a
successful in-order receive schedules a cumulative ACK.  Eleven module
tests cover ordering, retry, bounded state, profile selection, replay, and
control accounting.

The standard-replay path follows the public UALink 200G 1.0 semantics relevant
to our hypothesis \cite{ualink10}: a replay entry remains until cumulative ACK;
a request is not an implicit ACK; replay begins at the requested sequence and
includes subsequent retained payload flits; three request copies are generated;
and new payload admission stops while replay is active.  ACK and request bytes
share a serialized reverse link.  The simulator abstracts sequence-number wrap,
explicit-flit cadence, ambiguity handling, request-ignore timing, NOPs, exact
FEC interleave placement, and bit-level RS decoding so that the evaluation can
focus on replay exposure, feedback serialization, and FEC/exposure control.

Table~\ref{tab:protocol-map} separates the protocol properties required by our
hypothesis from abstractions that could change absolute timing.  This mapping
also separates UALink-style standard replay from NACK-only designs such as RIFL:
both retry at the link layer, but their retained state, feedback, and recovery
ordering differ.

\begin{table}[t]
\centering
\caption{Protocol-to-model mapping.}
\label{tab:protocol-map}
\resizebox{\columnwidth}{!}{%
\begin{tabular}{p{0.27\columnwidth}p{0.31\columnwidth}p{0.32\columnwidth}}
\toprule
Property & Modeled behavior & Relevance / abstraction \\
\midrule
TxReplay retention & Release only on cumulative ACK & Determines suffix state \\
Replay request & Three copies; no implicit ACK & Preserves requested start \\
Replay service & Requested sequence through retained tail & Core amplification effect \\
Admission & Block new payload during active replay & Couples replay to throughput \\
Reverse control & ACK/request share serializer & Exposes delayed feedback \\
FEC & Threshold correction, byte overhead, delay & No bit-level RS/miscorrection \\
Sequence control & In-order receive and retry limit & No wrap/ambiguity/NOP cadence \\
\bottomrule
\end{tabular}}
\end{table}

\subsection{Parameters and traffic}

Table~\ref{tab:params} lists the primary configuration.  GOOD-state corrupt
flits have at most one bad symbol; BAD-state flits have up to 24.  LIGHT and
STRONG profiles correct 4 and 15 symbols.  The 640-byte payload unit is chosen
to match the scale of a UALink-style data-link reliability unit and a
KP4-like RS-FEC codeword: 640~B plus 6.25\% modeled STRONG overhead is about
680~B \cite{ualink10,g7095}.  We use this scale as a modeling granularity for
UALink-style reliability units; other scale-up fabrics may choose different
flit formats.

The 14-ns offered period corresponds to 365.7~Gb/s of payload before FEC and
replay.  Every compared policy receives the same seeded error sequence.  Under
the primary GOOD/BAD probabilities, the stationary fraction of first-attempt
flits with any raw symbol error is about
$0.9524\cdot0.001+0.0476\cdot0.6\approx2.95\%$; the measured mean is 309.2 per
10,000 flits.  This controlled burst-stress workload makes replay amplification
observable at a 12.5-flit mean BAD duration.  Section~\ref{sec:validity}
describes the trace-calibration path for field BER and reliability targets.

\begin{table}[t]
\centering
\caption{Primary link-level configuration.}
\label{tab:params}
\resizebox{\columnwidth}{!}{%
\begin{tabular}{ll}
\toprule
Parameter & Value \\
\midrule
Payload flit / data rate & 640 B / 400 Gb/s \\
Forward / reverse rate & 400 / 400 Gb/s \\
Propagation; ACK/request base delay & 20 ns; 50 ns \\
TxReplay window; risk cap & 16; 8 flits \\
Request copies; retry limit & 3; 20 \\
FEC LOW & modeled 0 overhead, 0 ns, telemetry floor \\
FEC LIGHT & 2.5\%, 5 ns, 4 symbols \\
FEC STRONG & 6.25\%, 12 ns, 15 symbols \\
GOOD$\rightarrow$BAD / BAD$\rightarrow$GOOD & 0.004 / 0.08 per flit \\
Corruption emission GOOD / BAD & 0.001 / 0.6 \\
Offered period / flits per seed & 14 ns / 10,000 \\
\bottomrule
\end{tabular}}
\end{table}

\subsection{Parameter selection and data isolation}

We reserve seeds 1--10 for parameter selection.  All reported primary seeds
start at 101, robustness at 701, collective experiments at 1001 or 1201, and
the million-flit check at 2001.  Table~\ref{tab:training} shows the complete
two-stage training sweep.  First, increasing $\lambda$ trades more FEC for
fewer residual replays.  We choose 16 rather than 24 because their goodput is
within 0.3\%, while $\lambda=24$ nearly doubles redundancy.  Second, with
$\lambda=16$, the 8-flit risk budget dominates the larger candidates on
goodput, P99, replay, and FEC bytes.  No held-out or collective result is used
to change either parameter.

\begin{table}[t]
\centering
\caption{Training-only selection (10 seeds, 5,000 flits/seed).}
\label{tab:training}
\resizebox{\columnwidth}{!}{%
\begin{tabular}{llrrrr}
\toprule
Sweep & Value & Goodput & P99 & Replay & FEC \\
 & & (Gb/s) & ($\mu$s) & (flits) & (KB) \\
\midrule
$\lambda$ & 4  & 285.1 & 20.63 & 1463 & 29.6 \\
 & 8  & 294.4 & 17.60 & 1312 & 31.8 \\
 & 12 & 301.9 & 15.39 & 1200 & 34.8 \\
 & \textbf{16} & 315.9 & 11.72 & 976 & 116.1 \\
 & 24 & 315.0 & 11.56 & 737 & 224.2 \\
\midrule
$W_R$ & \textbf{8} & 336.0 & 6.72 & 583 & 108.5 \\
 & 10 & 326.6 & 8.99 & 855 & 115.0 \\
 & 12 & 322.4 & 10.04 & 918 & 115.3 \\
 & 14 & 315.9 & 11.72 & 976 & 116.1 \\
 & 16 & 310.9 & 13.13 & 996 & 116.3 \\
\bottomrule
\end{tabular}}
\end{table}

\subsection{Baselines and metrics}

We compare four policies over the same standard-replay state machine:

\begin{itemize}
  \item \textbf{Fixed strong FEC}: STRONG protects every attempt.
  \item \textbf{Post-decode adaptation}: an EWMA with hysteresis observes only
  decode failures and selects LOW/LIGHT/STRONG.  It represents the strongest
  reactive baseline we model.
  \item \textbf{Predictive FEC}: Eq.~\eqref{eq:cost} selects FEC from pre-FEC
  evidence but retains the 16-flit window.  This is the strict ablation.
  \item \textbf{PREFACE}: predictive FEC plus Eq.~\eqref{eq:window}.
\end{itemize}

Metrics are payload goodput, arrival-to-in-order-delivery P99 latency, replayed
payload flits, FEC redundancy bytes, and reverse-control bytes.  The primary
experiment uses 30 held-out seeds (101--130).  We report normal-approximation
95\% confidence intervals; comparisons use per-seed paired differences.  A
3$\times$3 robustness matrix uses 20 seeds for offered periods 13/14/16 ns and
BAD-to-GOOD probabilities 0.03/0.08/0.20.

\subsection{Modeled ring AllReduce}

To test whether link gains survive synchronization, we build a ring of 8, 16,
or 32 independent links.  A tensor is divided evenly among ranks and a
synchronous ring executes $2(N-1)$ reduce-scatter/all-gather steps, following
the standard bandwidth-optimal collective structure used in MPI and GPU
collective libraries \cite{thakur2005,patarasuk2009,nccl}.  Every step waits
for the slowest link, which exposes replay tail amplification.  Tensor sizes
are 1 and 8 MiB per rank.  The model captures the communication schedule over
replay-capable links while abstracting compute overlap, routing, switch
arbitration, and kernel launch costs in a full GPU/NIC software stack.

\subsection{Statistical procedure and reproducibility}

All policies in one comparison consume the same latent channel-state and
symbol-error stream, so we report per-seed paired deltas with two-sided 95\%
normal-approximation confidence intervals.  Training seeds are separated from
held-out reporting seeds, and the 3$\times$3 robustness grid is predefined.
The intervals quantify simulation-seed variability for this controlled mechanism
study.  The artifact records commands, seeds, raw aggregate CSVs, paired deltas,
and plotting scripts.

\section{Evaluation}

\subsection{Primary link-level result}

\begin{figure}[t]
  \centering
  \includegraphics[width=\columnwidth]{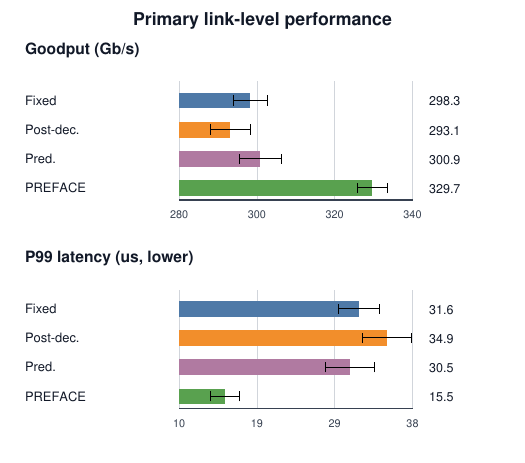}
  \caption{Held-out link-level means and 95\% confidence intervals (30 seeds).
  PREFACE improves both predictive FEC alone and the two replay baselines.}
  \label{fig:primary}
\end{figure}

Figure~\ref{fig:primary} and Table~\ref{tab:primary} show the primary result.
Relative to fixed strong FEC, PREFACE raises goodput from 298.32 to
329.71~Gb/s: a paired gain of $31.39\pm1.51$~Gb/s (10.52\%).  P99 falls from
31.56 to 15.54~$\mu$s, a reduction of $16.02\pm1.06$~$\mu$s (50.75\%).
PREFACE also removes $488.9\pm55.6$ replayed flits and spends
$231.8\pm1.5$~KB less FEC.  Reverse control increases by
$196.4\pm19.4$~KB because faster progress produces ACKs sooner, but total
FEC-plus-control still falls from 7.428 to 7.393~MB.  The modeled
reliability-traffic accounting therefore still favors PREFACE.

Against post-decode adaptation, PREFACE gains $36.60\pm1.75$~Gb/s (12.49\%),
lowers P99 by $19.40\pm1.39$~$\mu$s (55.51\%), and removes
$1245.5\pm77.0$ replayed flits (47.52\%).  It spends 127.4~KB more FEC but
sends 154.9~KB less reverse control.  The reactive baseline economizes on
coding only after uncorrectable frames have already triggered costly suffix
replay.

\begin{table}[t]
\centering
\caption{Primary means over 30 held-out seeds.}
\label{tab:primary}
\resizebox{\columnwidth}{!}{%
\begin{tabular}{lrrrr}
\toprule
Policy & Goodput & P99 & Replay & FEC \\
 & (Gb/s) & ($\mu$s) & (flits) & (KB) \\
\midrule
Fixed strong FEC & 298.32 & 31.56 & 1864.1 & 458.4 \\
Post-decode adapt. & 293.11 & 34.94 & 2620.7 & 99.2 \\
Predictive FEC & 300.88 & 30.50 & 2312.4 & 244.4 \\
PREFACE & \textbf{329.71} & \textbf{15.54} & \textbf{1375.2} & \textbf{226.6} \\
\bottomrule
\end{tabular}}
\end{table}

Table~\ref{tab:paired} reports paired effects rather than only policy means.
Every interval for the three performance metrics excludes zero.  The overhead
rows expose the different tradeoffs: PREFACE saves both FEC and total modeled
reliability bytes versus fixed FEC; versus the low-coding reactive policy, it
spends coding bytes to avoid a much larger replay and tail penalty; versus
predictive FEC, it saves both FEC and reverse control because the only changed
actuator is exposure.  This last column is particularly important because it
rules out a hidden ``more redundancy'' explanation for the ablation gain.

\begin{table}[t]
\centering
\caption{Paired PREFACE deltas with 95\% CI (30 held-out seeds).}
\label{tab:paired}
\resizebox{\columnwidth}{!}{%
\begin{tabular}{lrrr}
\toprule
Metric & vs. fixed & vs. post-decode & vs. predictive \\
\midrule
Goodput (Gb/s) & $+31.39\pm1.51$ & $+36.60\pm1.75$ & $+28.84\pm1.96$ \\
P99 ($\mu$s) & $-16.02\pm1.06$ & $-19.40\pm1.39$ & $-14.96\pm1.36$ \\
Replay (flits) & $-489\pm56$ & $-1246\pm77$ & $-937\pm72$ \\
FEC (KB) & $-231.8\pm1.5$ & $+127.4\pm2.4$ & $-17.8\pm1.6$ \\
Control (KB) & $+196.4\pm19.4$ & $-154.9\pm18.9$ & $-23.0\pm11.7$ \\
\bottomrule
\end{tabular}}
\end{table}

\subsection{Does joint exposure control matter?}

The predictive-FEC ablation holds the error trace, posterior, cost function,
coding choice, and replay machine constant; only $W_t$ remains 16.  PREFACE adds
$28.84\pm1.96$~Gb/s, reduces P99 by $14.96\pm1.36$~$\mu$s, and removes
$937.2\pm71.6$ replayed flits.  FEC redundancy also decreases by
$17.80\pm1.64$~KB and reverse control by $23.0\pm11.7$~KB.  This is the central
causal result: holding predictive FEC fixed, risk-capping reduces the amount of
healthy-looking suffix traffic invalidated by a residual failure.  The large
difference also explains why selecting the
risk window after reverse-control serialization was added was necessary: a
window trained under an earlier timing model would not represent the finalized
system.

\subsection{Robustness across load and burst duration}

\begin{figure}[t]
  \centering
  \includegraphics[width=\columnwidth]{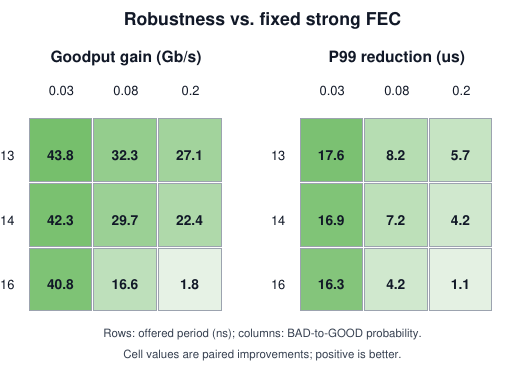}
  \caption{Paired PREFACE improvement over fixed strong FEC.  Positive is
  better; corresponding paired intervals exclude zero.}
  \label{fig:robustness}
\end{figure}

PREFACE significantly improves goodput, P99, and replay count against both
fixed strong FEC and post-decode adaptation in all 9/9 robustness scenarios.
It also uses less FEC than fixed strong FEC at every point.  Figure
\ref{fig:robustness} shows that gains grow when bursts persist or offered load
is high enough for delayed feedback to accumulate a suffix.  At the hardest
13-ns/0.03 point, PREFACE gains $43.78\pm3.29$~Gb/s and lowers P99 by
$17.65\pm2.40$~$\mu$s versus fixed FEC.  The same point gains
$45.80\pm3.58$~Gb/s over post-decode adaptation.

The useful effect does narrow under light load and short bursts, as predicted
by Eq.~\eqref{eq:exposure-bound}.  At 16 ns and BAD-to-GOOD 0.20, the fixed-FEC
comparison is only $1.81\pm1.29$~Gb/s goodput and
$1.09\pm0.32$~$\mu$s P99, although both remain significant.  Replay falls by
$55.4\pm23.2$ flits.  This boundary matches the mechanism: PREFACE removes
replay exposure, so the benefit naturally approaches zero when the offered
stream rarely fills the healthy window or BAD episodes end before feedback
accumulates a suffix.

\subsection{Overhead and control-path interpretation}

PREFACE occupies a middle point between the coding baselines.  In the primary
experiment it uses 226.6~KB of FEC per seed, 50.6\% less than always-strong FEC
and 128.4\% more than post-decode adaptation.  The relevant cost also includes
replay: the latter baseline triggers 90.6\% more replayed flits than
PREFACE.  Reverse-control bytes are 2.2\% lower than post-decode adaptation and
2.8\% higher than fixed FEC.  These magnitudes show why evaluating only coding
overhead, or assuming instantaneous ACK/NACK delivery, would reverse some
design conclusions.

The flight cap uses the existing replay-request format in our model.  Its
wire-visible consequence is indirect: fewer suffix replays and fewer repeated
requests.  A real profile negotiation mechanism may add small control messages
or reserve profile epochs; deployment studies should account for those bytes
explicitly.

\subsection{Ring AllReduce}

\begin{figure}[t]
  \centering
  \includegraphics[width=\columnwidth]{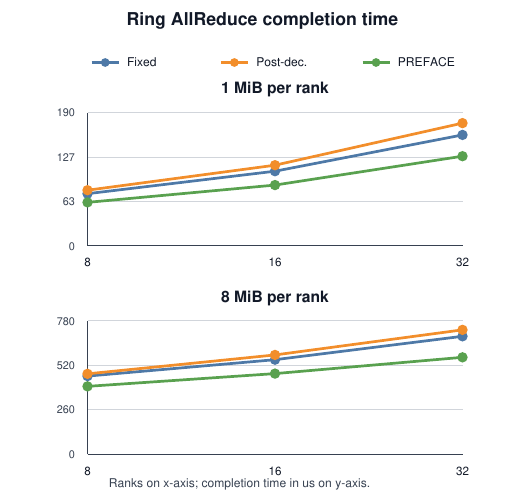}
  \caption{Modeled synchronous ring AllReduce completion time.  PREFACE is
  significantly faster than both baselines in all six configurations.}
  \label{fig:collective}
\end{figure}

Figure~\ref{fig:collective} shows that link-level gains survive a synchronized
collective schedule.  Relative to post-decode adaptation, PREFACE reduces
completion time in all six configurations: 21.9--27.0\% for 1 MiB and
15.6--22.1\% for 8 MiB.  Relative to fixed strong FEC, reductions are
16.6--19.2\% for 1 MiB and 13.1--17.8\% for 8 MiB.  Every paired interval
excludes zero.  The rank trend is also favorable: for 1 MiB, the fixed-FEC
reduction rises from $12.37\pm1.48$~$\mu$s at 8 ranks to
$30.30\pm1.84$~$\mu$s at 32 ranks because more synchronized links create more
opportunities for a replay tail to determine the step maximum.

At 32 ranks and 8 MiB, completion falls from 689.5 to 566.6~$\mu$s versus fixed
FEC (17.83\%) and from 727.2 to 566.6~$\mu$s versus post-decode adaptation
(22.09\%).  Paired reductions are $122.92\pm4.02$ and
$160.60\pm8.03$~$\mu$s.  Replay falls by 32,261 and 89,680 flits,
respectively.  This communication-schedule result shows that independent link
benefits survive slowest-link synchronization, while full end-to-end training
speed depends on compute overlap, software, routing, and switch effects.

\subsection{Long-run delivery invariant}

Three additional seeds each offer one million flits.  All three million flits
(1.92~GB payload) are delivered in order with zero terminal failures, despite a
mean of 29,586 initially corrupted and 130,060 replayed flits per seed.  The
online transition counters recover the injected 0.004/0.08 probabilities,
which checks that the shorter flight budget preserves the injected error-process
statistics.
The check targets delivery-order conservation on the modeled detected-error
path, which uses perfect detection, threshold-bounded correction, and eventual
success within 20 retries.  A later FIT/SDC study should extend the fault model
with CRC collisions, decoder miscorrection, analog behavior, and permanent or
common-mode faults.

\section{Related Work}

\textbf{Link-local retransmission.}  RIFL develops a low-latency in-band,
NACK-only hop-by-hop retransmission protocol and demonstrates a 112-Gb/s FPGA
implementation \cite{rifl}.  Its continuous idle frames, bidirectional-error
handling, verification code, and replay state illustrate a different
link-retry design point from UALink-style retained standard replay.  PREFACE is
complementary: it controls coding and exposure before an existing replay machine
activates.  Scale Up Ethernet likewise identifies link-layer retry as recovery
beyond FEC \cite{sue}.

\textbf{Adaptive FEC.}  Adaptive coding has a long history in wireless and
streaming networks; for example, burst-loss measurement can drive streaming
FEC redundancy \cite{adaptivefec}.  PREFACE builds on adaptive coding and
Gilbert--Elliott-style prediction, and adds the scale-up-specific coupling
between a pre-FEC posterior and the standard-replay exposure window.  The
fixed-window predictor ablation isolates the value of that coupling.

\textbf{Combining coding and retransmission.}  Earlier systems demonstrate that
coding and ARQ are complementary but operate at very different scales.
Maelstrom aggregates wide-area traffic and applies layered interleaving so that
FEC recovery can happen before a long RTT dominates \cite{maelstrom}; Ricochet exchanges
receiver-generated repair information for time-critical multicast
\cite{ricochet}.  WiLDNet explicitly studies the throughput/delay/error tradeoff
between bulk link retransmissions and FEC on long-distance wireless links
\cite{wildnet}.  FlEC exposes multiple coding/retransmission choices to QUIC
applications \cite{flec}.  These systems decide how to recover packet loss or
where to place redundancy.  PREFACE instead observes corrected physical errors
at a sub-packet scale and controls how much standard-replay state becomes
conditionally exposed before a request returns.  The relationship is conceptual:
all of these systems value proactive and reactive recovery, while PREFACE applies
that principle to flit-scale replay exposure.

\textbf{Scale-up reliability.}  UALink specifies CRC/FEC/replay machinery
\cite{ualink10}; RXL proposes implicit sequence numbers and end-to-end integrity
for scaling CXL-like chip networks \cite{rxl}.  These works define or extend the
correctness path.  PREFACE instead optimizes when and how much traffic reaches
that path without changing its correctness semantics.

\section{Discussion}

\subsection{Alternatives and design rationale}

\textbf{Always use the risk window.}  A static 8-flit TxReplay window would
bound suffix exposure without a predictor, but it would also impose the
underfill term $u(W)$ during the overwhelmingly common GOOD state.  PREFACE
keeps 16 entries whenever the selected code is LOW or LIGHT and contracts only
under the same evidence that justifies STRONG FEC.  Its window cap is therefore
a transient risk budget layered on top of ordinary credit sizing.  A
static-small-window baseline would answer a different question---how much
throughput to sacrifice universally for a replay bound---and should be included
in hardware follow-up once the actual pipeline depth is known.

\textbf{Encode harder but never cap flight.}  This is precisely the predictive
FEC ablation.  It avoids reacting after decode failure, yet its 300.88~Gb/s
goodput remains close to fixed FEC because a residual event can still couple to
the full outstanding suffix.  Increasing $\lambda$ raises coding pressure while
leaving the exposure term unchanged: the training sweep shows that
$\lambda=24$ roughly doubles FEC bytes while goodput changes by less than
1~Gb/s.  Available profiles can reduce residual probability; flight control
reduces the conditional cost of the residual events that remain.

\textbf{Replay only the damaged flit.}  Selective replay could retransmit only
the failed flit or receiver-reported holes, directly reducing suffix
amplification.  It is a protocol redesign: the receiver must retain and validate
out-of-order flits, requests need finer-grained metadata or bitmaps, and
verification must cover hole tracking, duplicate replays, ACK races, and buffer
reclamation.  PREFACE preserves standard replay and reduces replay probability
and exposure before recovery activates.  The ideas are complementary; future
work should compare selective replay, standard replay plus PREFACE, and their
combination under the same burst traces and hardware cost model.

\textbf{FEC profile transition cost.}  The simulator abstracts profile changes
as immediate decisions over a multi-profile FEC interface.  Real links must
synchronize which profile protects each flit or epoch, protect profile metadata,
and handle replay across profile boundaries.  Header IDs, epoch switching, and
rate-compatible codes are possible paths, but they add control bytes, transition
latency, decoder complexity, and verification cost.  A conservative design can
force STRONG FEC when state is uncertain.  These costs affect net performance;
standard replay remains the detected-error correctness backstop.

\textbf{Use a learned predictor.}  A recurrent or transformer model could
consume richer PHY telemetry.  The two-state filter used here keeps the
observation, transition, cost, and safety behavior inspectable, and its handful
of state variables is compatible with per-flit control.  Trace-calibrated
learned prediction is best viewed as a replaceable front end to the joint
actuation policy, with explicit latency, area, and out-of-distribution fallback
costs.

The measured results are consistent with the exposure model.  The fixed-window
predictive ablation sees exactly the same error samples and coding decisions
yet retains substantially more replay and tail latency.  Gains increase with
load, burst persistence, and rank count---the three factors that increase the
number or impact of suffix replays.  PREFACE also lowers FEC relative to fixed
strong FEC and replay relative to post-decode adaptation, connecting the
measured effect to Eq.~\eqref{eq:joint-cost} rather than to a single
``more-coding'' or ``less-coding'' explanation.

\subsection{Deployment path and validation needs}
\label{sec:validity}

The main calibration need is trace quality.  The Gilbert--Elliott parameters and
symbol magnitudes are synthetic; field validation should collect pre-FEC error
traces across cables, connectors, temperatures, lane training, and aging, then
replay those traces or fit a richer semi-Markov model.  A bit-accurate RS
decoder is also needed to study miscorrection and post-FEC undetected-error
rates.

The protocol model captures the UALink 200G 1.0-style standard-replay
interactions needed by the hypothesis and keeps timing-sensitive details as
explicit extension points.  Later or gated UALink revisions, including 2.0
materials \cite{ualink20}, are natural follow-on validation targets.  ACK cadence
currently uses one 640-byte control flit per payload flit; future models should
add cumulative ACK coalescing and reverse payload contention.

Implementation confidence comes from deterministic seeded traces, paired policy
runs, explicit counters for delivered, corrupted, corrected, replayed, and
terminal frames, and 11 module tests.  The long-run invariant check exercises
conservation at a larger event count, while the reported confidence intervals
summarize held-out seed variability on the predefined robustness grid.

Two deployment details deserve explicit engineering.  First, profile changes are
immediate in the simulator; hardware needs safe codeword transition epochs,
protected profile metadata, fixed-point posterior arithmetic, timing closure,
and area/power accounting.  Second, LOW needs useful early telemetry, either from
a monitor code that reports corrected symbols or from equivalent PHY-side
pre-FEC counters.  The ring experiment captures collective synchronization over
modeled links; trace-driven collective simulation and FPGA or SerDes testbeds
are the next steps for application-speed and reliability evaluation.

\section{Future Work}

Four directions are important for turning PREFACE from a controlled mechanism
study into a deployable scale-up reliability mechanism.

\subsection{Trace-calibrated channel and correctness models}

The first priority is to replace the synthetic Gilbert--Elliott input with
pre-FEC traces from high-speed electrical links.  A useful campaign should
span cable length, connector quality, temperature, equalizer settings, lane
training, and aging.  Besides corrected-symbol counts, traces should retain
time, lane identity, codeword position, decoder confidence when available, and
whether an event crossed the correction threshold.  We can then compare the
two-state filter against semi-Markov, change-point, and lightweight learned
predictors using identical train/test splits.  The key metric is delivered cost
after inference delay and fallback behavior are included, rather than prediction
accuracy alone.

A bit-accurate FEC/CRC path is needed in parallel.  It should model
interleaving, decoder miscorrection, CRC coverage, and correlated multi-lane or
common-mode faults.  That model would separate three quantities currently
collapsed by a threshold abstraction: corrected errors that inform prediction,
detected residual errors that trigger replay, and undetected corruptions that
threaten integrity.  The viability condition is clear: early telemetry should
predict risky intervals while preserving undetected-error probability and
profile-transition safety.  This work would support SDC/FIT analysis by
replacing the current invariant check with a richer correctness model.

\subsection{Protocol-complete feedback and multi-link behavior}

The next simulator revision should implement cumulative ACK coalescing,
explicit-flit cadence, sequence wrap and ambiguity handling, request-ignore
timers, NOP behavior, and reverse payload contention.  It should also cross-check
the modeled state machine against an independently available specification or
interoperable implementation.  These additions can change the detection-to-
request interval in Eq.~\eqref{eq:exposure-bound}, and therefore may move the
optimal absolute risk budget.  A useful stress test is whether joint control
remains beneficial after ACK aggregation reduces feedback delay.

The risk budget itself should eventually be adaptive.  PREFACE fixes $W_R=8$
after offline training, but a production controller could compute $W_t$ from the
online posterior, estimated GOOD/BAD transitions, residual cost, offered load,
and feedback delay.  This would avoid over-conservative caps on benign links
while tightening flight under frequent or persistent bursts.

Scale-up packages also contain multiple lanes and links whose faults need not
be independent.  Future experiments should introduce lane striping, shared
power/clock disturbances, link bonding, and failover.  Controller state could
be maintained per lane, per link, or hierarchically.  Per-lane state reacts
precisely but costs more; a shared posterior is smaller but may overprotect
healthy lanes.  This creates a new question beyond the current paper: whether
exposure should be capped locally or coordinated across the striped transfer to
prevent one risky lane from becoming the collective tail.

\subsection{Hardware prototype and safe deployment}

A hardware prototype should separate fast-path inference from slow-path model
maintenance.  The per-flit path needs a posterior register, a small table or
fixed-point evaluator for three residual costs, a minimum comparator, and an
admission-limit register.  Transition-probability estimation, diagnostics, and
policy updates can run at a slower epoch.  Profile changes must be associated
with an unambiguous codeword boundary; an epoch tag, ordered management message,
or prearranged profile schedule are possible choices, but selecting among them
requires implementation details below the public model used here.

The safest first prototype would keep strong FEC as a boot and fault profile,
enable PREFACE only after both ends acknowledge telemetry, and revert on
posterior saturation, telemetry loss, or control timeout.  FPGA evaluation
should report inference latency, maximum clock rate, logic and memory area,
admission-gate critical path, energy per delivered bit, and transition-control
bytes.  Fault injection should test false positives, false negatives, permanent
faults, and mid-transition resets.  A useful success criterion should combine
ns-3 goodput agreement with preservation of the baseline replay invariant and
timing closure at the target flit rate.

\subsection{Full-stack collective evaluation}

The final track is to connect link behavior to a GPU/NIC communication stack.
Trace-driven simulation should combine measured replay episodes with collective
algorithms, routing, switch arbitration, compute overlap, and kernel launch
costs.  Beyond ring AllReduce, the workload set should cover trees, all-gather,
all-to-all, point-to-point transfers, and mixtures of latency-sensitive control
messages with bulk tensors.  Model outputs should include job completion and
iteration jitter in addition to average collective time, because one exposed
suffix on a synchronized path can dominate a step.

There is also an opportunity for cross-layer control.  The link can export a
coarse risk epoch without exposing raw PHY telemetry; a collective scheduler
could temporarily avoid a degraded path, change chunking, or delay a noncritical
transfer.  Conversely, the communication layer could identify synchronization-
critical flits whose replay exposure deserves a smaller budget.  Such coupling
must be compared against link-local PREFACE to determine whether its additional
interface and policy complexity produce gain beyond the simple local design.

Table~\ref{tab:roadmap} summarizes a staged agenda.  Each phase has an explicit
artifact and a decision criterion for judging whether the mechanism remains
compelling.

\begin{table}[t]
\centering
\caption{Proposed validation roadmap.}
\label{tab:roadmap}
\resizebox{\columnwidth}{!}{%
\begin{tabular}{p{0.12\columnwidth}p{0.31\columnwidth}p{0.43\columnwidth}}
\toprule
Phase & Deliverable & Key validation question \\
\midrule
Trace & Public/anonymized pre-FEC burst corpus & Does early telemetry predict risky intervals? \\
Model & Bit-accurate FEC/CRC and complete replay & Do joint gains survive accurate feedback? \\
HW & FPGA/SerDes controller with fault injection & Do timing/safety costs stay below saved replay work? \\
Stack & Trace-driven collectives and application runs & Do link gains survive software/topology effects? \\
\bottomrule
\end{tabular}}
\end{table}

These phases also motivate shared artifacts: a pre-FEC error-trace format, a
protocol-semantic replay test suite, and common reporting of FEC, replay,
control, tail latency, and delivered-integrity costs.  Such artifacts would let
future mechanisms be compared on more than a single BER or throughput number.

\balance

\section{Conclusion}

Delayed standard replay couples error correction to the amount of traffic in
flight.  PREFACE makes that coupling explicit: a pre-FEC burst posterior selects
both coding strength and a risk-capped exposure window, while standard replay
remains the correctness backstop.  In the final serialized-control ns-3 model,
PREFACE improves goodput and tail latency over fixed strong FEC and post-decode
adaptation, and a strict ablation shows that joint flight control contributes
substantial gain beyond predictive FEC.  Robustness and collective experiments
identify both the useful region and the saturation points.  The evidence
supports PREFACE as a promising scale-up link-control design, with hardware and
trace calibration as the essential next validation step.

\bibliographystyle{ACM-Reference-Format}
\bibliography{references}

\end{document}